\documentclass[11pt]{article}

\usepackage{fullpage}
\usepackage{setspace}
\usepackage{parskip}
\usepackage{titlesec}
\usepackage[section]{placeins}
\usepackage{xcolor,colortbl}
\usepackage{breakcites}
\usepackage{lineno}

\usepackage[utf8]{inputenc}
\usepackage[english]{babel}

\usepackage{times}

\PassOptionsToPackage{hyphens}{url}
\usepackage[colorlinks = true,
            linkcolor = blue,
            urlcolor  = blue,
            citecolor = blue,
            anchorcolor = blue]{hyperref}
\usepackage{etoolbox}

\renewenvironment{abstract}
  {{\bfseries\noindent{\abstractname}\par\nobreak}\footnotesize}
  {\bigskip}

\titlespacing{\section}{0pt}{*3}{*1}
\titlespacing{\subsection}{0pt}{*2}{*0.5}
\titlespacing{\subsubsection}{0pt}{*1.5}{0pt}

\usepackage{authblk}

\usepackage{graphicx}
\usepackage[space]{grffile}
\usepackage{latexsym}
\usepackage{textcomp}
\usepackage{longtable}
\usepackage{tabulary}
\usepackage{booktabs,array,multirow}
\usepackage{amsfonts,amsmath,amssymb}
\providecommand\citet{\cite}
\providecommand\citep{\cite}

\newif\iflatexml\latexmlfalse

\usepackage{float}

\usepackage{tabularx}
\usepackage{multirow}
\usepackage{subfig}
\usepackage[export]{adjustbox}
\usepackage{amsbsy}
\usepackage{manyfoot}
\usepackage{algorithm}
\usepackage{algorithmicx}
\usepackage{algpseudocode}
\usepackage{listings}
\usepackage{url}

\usepackage[margin=0.85in]{geometry}

\begin{document}

\title{Multilayer Quantile Graph for Multivariate Time Series Analysis and Dimensionality Reduction}

\author[1]{Vanessa Freitas Silva}
\author[2]{Maria Eduarda Silva}
\author[1]{Pedro Ribeiro}
\author[1]{Fernando Silva}

\affil[1]{CRACS-INESC TEC, Faculdade de Ciências, Universidade do Porto}
\affil[2]{LIAAD-INESC TEC, Faculdade de Economia, Universidade do Porto}

\vspace{-1em}

\date{}

\begingroup
\let\center\flushleft
\let\endcenter\endflushleft
\maketitle
\endgroup

\selectlanguage{english}
\begin{abstract}
In recent years, there has been a surge in the prevalence of high- and multi-dimensional temporal data across various scientific disciplines. These datasets are characterized by their vast size and challenging potential for analysis. Such data typically exhibit serial and cross-dependency and possess high dimensionality, thereby introducing additional complexities to conventional time series analysis methods. To address these challenges, a recent and complementary approach has emerged, known as network-based analysis methods for multivariate time series. In univariate settings, Quantile Graphs have been employed to capture temporal transition properties and reduce data dimensionality by mapping observations to a smaller set of sample quantiles. 

To confront the increasingly prominent issue of high dimensionality, we propose an extension of Quantile Graphs into a multivariate variant, which we term "Multilayer Quantile Graphs". In this innovative mapping, each time series is transformed into a Quantile Graph, and inter-layer connections are established to link contemporaneous quantiles of pairwise series. This enables the analysis of dynamic transitions across multiple dimensions. In this study, we demonstrate the effectiveness of this new mapping using a synthetic multivariate time series dataset. We delve into the resulting network's topological structures, extract network features, and employ these features for original dataset analysis. Furthermore, we compare our results with a recent method from the literature. The resulting multilayer network offers a significant reduction in the dimensionality of the original data while capturing serial and cross-dimensional transitions. This approach facilitates the characterization and analysis of large multivariate time series datasets through network analysis techniques.

\end{abstract}

\par\null

\section{Introduction}\label{sec1}

In recent years, the prevalence of multidimensional data has surged across various research fields thanks to technological advancements that have enabled the generation of vast datasets using advanced sensing technologies. While time series analysis, a well-established field~\cite{Sumway2017, Stoffernonlinear}, has traditionally focused on the analysis of time-indexed univariate data, the contemporary data landscape now encompasses high-dimensional, multivariate, and panel time series data collected concurrently from numerous sensors. Existing methods for analyzing such data are often constrained, designed for specific domains, and rely on various assumptions, leaving many unresolved challenges and hindering broader applications (see~\citet{wei2018multivariate} for more details). 
For instance, the high dimensionality of the data imposes limitations on computational and memory capacities, rendering the application of conventional methods difficult and often impractical. 

Efforts to confront the challenges posed by high-dimensional temporal data have emerged within the realms of data mining, machine learning, and network science. Specifically, network science offers a vast array of both elementary and complex topological features for characterizing various properties of network structures~\cite{Barabasi2016, Costa2007}. 
Recent developments have introduced advanced graph structures, known as multilayer networks, which facilitate the modeling of multidimensional data without sacrificing essential properties, including both intra-dimensional and inter-dimensional connections. Multilayer networks are intricate structures capable of establishing internal connections within the same layer/graph and external connections between different layers/graphs. Despite being a relatively recent addition to the field of network science, well-established methods, and methodologies can be readily extended and adapted to this innovative concept~\cite{kivela2014multilayer}.

In this study, our primary focus is on a recent multilayer network-based approach for analyzing and representing multivariate time series data. This approach is relatively novel, especially in the context of constructing multilayer networks. Traditional approaches often involve simplifying multivariate time series into single-layer networks, which can lead to the loss of valuable data information crucial for comprehensive analysis. Furthermore, existing methods that map multivariate time series to multilayer networks have raised questions, as indicated in prior research~\cite{vanessa2020,vanessa2023}. For instance, the multiplex visibility graphs~\cite{lacasa2015network,eroglu2018multiplex} only establish external connections between the same node across different adjacent layers, potentially overlooking direct external connections between different nodes. Another recent example is the Multilayer Horizontal Visibility Graphs~\cite{vanessa2023}, which can be computationally intensive and impractical for handling large datasets.

In the realm of univariate time series analysis, Quantile Graphs~\cite{Campanharo2011} have proven effective in mapping transition properties, offering the advantage of reducing data dimensionality. Given these properties and considering the aforementioned challenges, our work introduces a novel method for mapping multivariate time series data. This method extends the concept of Quantile Graphs to a multivariate context, resulting in what we term the "Multilayer Quantile Graph". The process entails establishing cross-dimensional connections between contemporaneous data quantile samples among the univariate components of the time series dataset. While connections between lagged data are also possible, in this work, we primarily focus on contemporary connections to introduce the method.

\begin{figure*}
    \centering
    \includegraphics[width=0.9\textwidth]{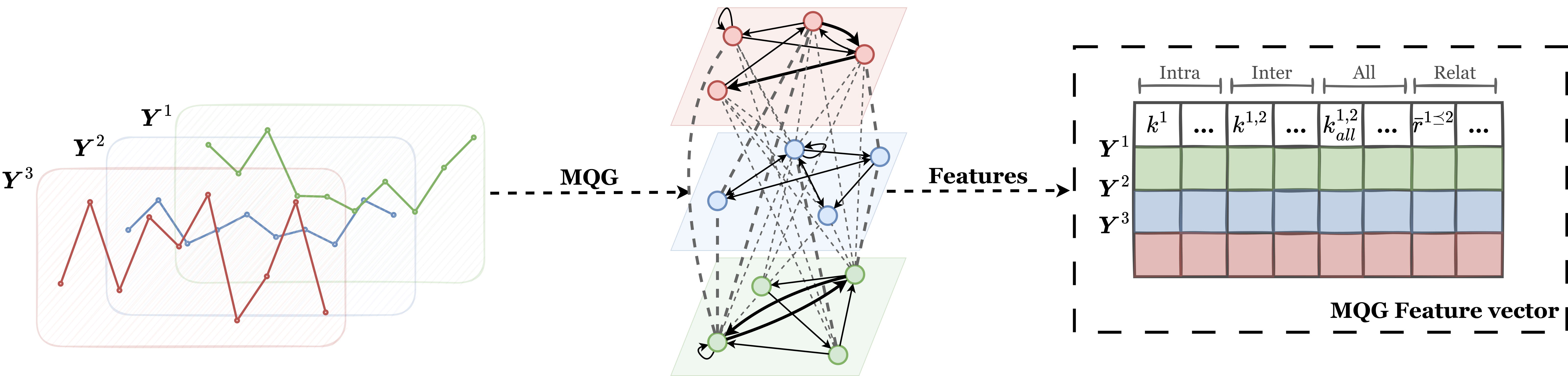}
    \caption{Schematic diagram of the multilayer network-based approach to multivariate time series reducing and mining.}
    \label{diagrm}
\end{figure*}

Our primary objective in this study is twofold: first, to introduce a new multivariate mapping method within the taxonomy of time series mappings~\cite{vanessa2023}, and second, to advance the representation of multivariate time series data using multilayer time series networks. To accomplish this, we evaluate the proposed mapping approach using a synthetic multivariate time series generated from a selected set of multivariate time series models. We analyze high-level topological properties proposed in~\cite{vanessa2023} through features extracted from the resulting Multilayer Quantile Graphs and use these properties to assess and analyze the mapping method. Additionally, we compare our results with a similar method, the Multilayer Horizontal Visibility Graph, which we introduced in our prior work~\cite{vanessa2023}. Figure~\ref{diagrm} illustrates the methodology employed in this study.

In summary, our contributions to this work can be summarized as follows: 
\begin{itemize}
    \item We enhance the interpretation of both univariate and multivariate time series mapping to develop a new and more efficient mapping method.
    
    \item We employ recently proposed high-level topological features~\cite{vanessa2023} for mapping analysis and evaluation, comparing our method with a recent approach from the literature.

    \item We show that our proposed mapping is computationally more efficient.
\end{itemize}

The remainder of the paper is organized as follows:

We start in Section~\ref{sec2} by introducing the necessary background and notation to facilitate the understanding of the subsequent sections.
In Section~\ref{sec3}, we provide a concise overview of the principal existing multivariate time series mapping approaches. 
Section~\ref{sec4} proposes a novel multivariate time series mapping method. 
In Section~\ref{sec5}, we elucidate our evaluation methodology and describe the experiments conducted.
Lastly, in Section~\ref{sec6}, we offer our conclusions, provide insights, and outline future research.

\section{Preliminaries}\label{sec2}
In this section, we introduce the necessary background and notation on multivariate time series and multilayer networks.

\subsection{Multivariate Time Series}
\label{subsec:mts}
We can think of time series data as collections of observations indexed by time. 
Formally, we can define a \textit{Univariate Time Series} (UTS) as a sequence of (scalar) observations time-indexed usually denoted by $\{{Y}_t\}_{t=1}^{T},$ and a \textit{Multivariate Time Series} (MTS) as a vector of $m$ observations obtained at each time $t$, i.e. $\boldsymbol{Y}_t = [Y_{1,t}, Y_{2,t}, \ldots, Y_{m,t}]^{\prime},$ where $\prime$ represents the transpose. 
 We denote an MTS by $\boldsymbol{Y}=\{\boldsymbol{Y}_t\}_{t=1}^{T}$ and the UTS components of the MTS $\boldsymbol{Y}$ by $\boldsymbol{Y}^{\alpha}=[Y_{\alpha,1}, Y_{\alpha,2}, \ldots, Y_{\alpha, T}]$ with $\alpha=1, \ldots,m,$ thus, we can denote a MTS data by its components, $\boldsymbol{Y}=\{Y^{\alpha}\}_{\alpha=1}^{m}.$ 

UTS is ordered in time and usually presents serial correlation as opposed to a random sample. MTS presents not only serial correlation within each UTS component, $\boldsymbol{Y}^{\alpha},$ but also (\textit{contemporaneous} and \textit{lagged}) correlation between the different UTS components, $\boldsymbol{Y}^{\alpha}$ and $\boldsymbol{Y}^{\beta}, \alpha \neq \beta.$ 
Thus, analyzing MTS depends on key dependence measures such as the \textit{autocorrelation function} (ACF), which measures the linear predictability of a UTS, and the \textit{cross-correlation function} (CCF), which measures the correlation between any two different UTS components of the MTS. 
The theory of UTS analysis is mature and solid, and although the methods and statistical models extend naturally to the multivariate case new issues and new concepts inevitably arise~\cite{wei2018multivariate}. 
An adequate MTS analysis requires advanced tools, methods, and models for mining information from multiple components that present temporal and cross-sectional correlations and impose high-dimensionality issues. 

\subsection{Multilayer Networks}
\label{subsec:mnet}

An alternative time series analysis approach is to map univariate and multivariate time series data to a network representation and use network science methodologies to analyze the original time series. 
Simply, a \textit{network} (or \textit{graph}) is a mathematical structure, $G=(V, E),$ that represents a set of elements by nodes, $V,$ and the connections between elements by a set of edges, $E.$ 

\begin{figure*}[htb!]
    \centering
    \includegraphics[width=0.8\textwidth]{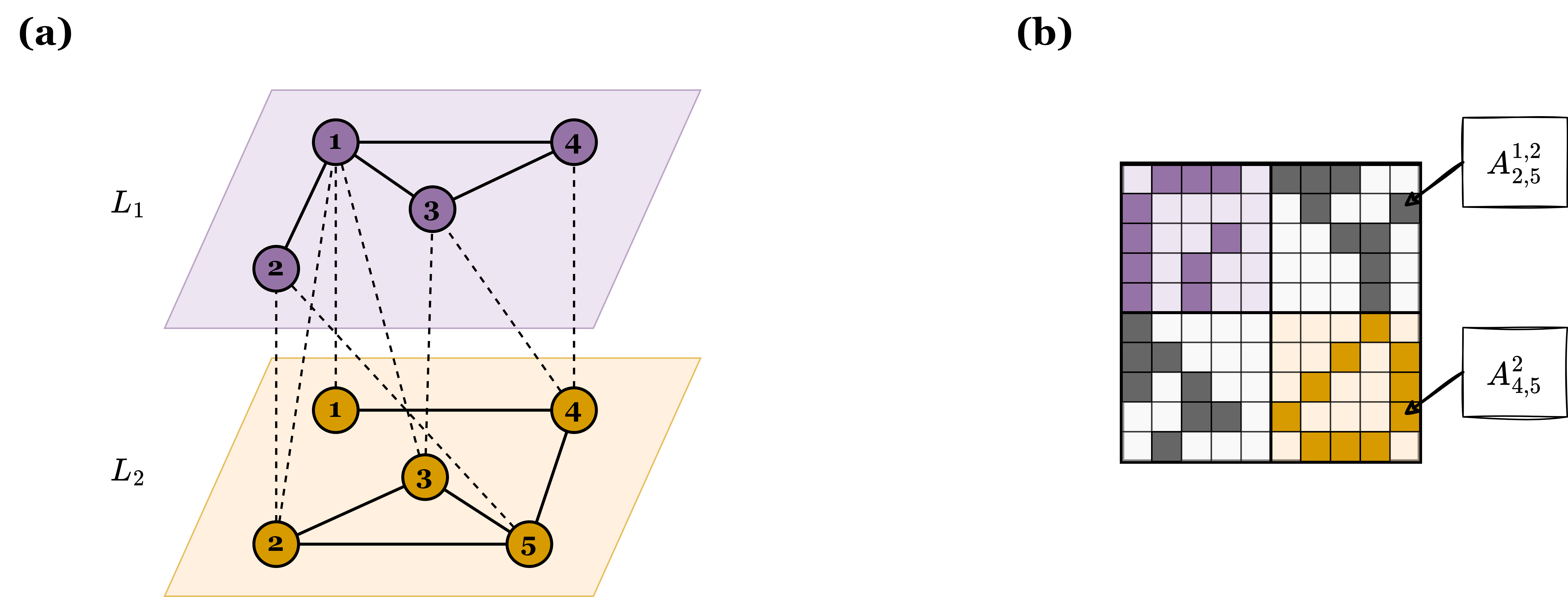}
    \caption[Illustration of the multilayer network and supra-adjacency matrices.]{An illustrative example of (a) a toy multilayer network with five entities $V= \{1,2,3,4,5\}$ and two elementary-layers $L_1$ and $L_2$, and (b) the corresponding supra-adjacency matrix. The solid lines (colored blocks) represent intra-layer edges and the dashed lines (gray blocks) represent inter-layer edges. 
    \textit{Source:} Modified from~\cite{vanessa2020}}
    \label{fig:mnet}
\end{figure*}

A \textit{Multilayer Network} (MNet) is a more general and complete definition of a network that can model several types of connections between elements of the same and different systems. 
Formally, a MNet is defined by $M = (V_M, E_M, V, \boldsymbol{L})$~\cite{kivela2014multilayer}, 
where 
    $V$ and $\boldsymbol{L}$ represent the sets of entities and layers, respectively, and 
    $V_M$ represents the set of node-layer combinations, $V_M \subseteq V \times L_1 \times \ldots \times L_m,$ in which a node is present in the corresponding elementary layer $L_{\alpha} \in \boldsymbol{L}.$    
    And $E_M \subseteq V_M \times V_M$ represents the set of edges (pairs of possible combinations of nodes and elementary layers), we call \textit{intra-layer edges} to the connections between nodes of the same layer, $(v_i^{\alpha}, v_j^{\alpha}),$ and \textit{inter-layer edges} to the connections between nodes of different layers, $(v_i^{\alpha}, v_j^{\beta})$ with $\alpha \neq \beta.$ 
A MNet can be represented by an \textit{adjacency tensor} of order $4$, $\pmb{\mathcal{A}},$  with tensor element $\mathcal{A}_{i,j,\alpha,\beta} = 1$ if $(v_i^{\alpha}, v_j^{\beta}) \in E_M$ and is $0$ otherwise~\cite{kivela2014multilayer}. 
Another representation is flattening $\pmb{\mathcal{A}}$ into a \textit{supra-adjacency matrix}, $\boldsymbol{A}$, where intra-layer edges are associated with diagonal element blocks and inter-layer edges with off-diagonal element blocks~\cite{vanessa2023}. 
So, we can infer the following types of subgraphs:
\newpage
\begin{itemize}
    \item \textit{intra-layer graphs}, $G^{\alpha}$, formed by the diagonal element blocks, $\left[ \begin{smallmatrix}   \boldsymbol{A}^{\alpha} & \boldsymbol{0}\\ \boldsymbol{0}  & \boldsymbol{0} &  \end{smallmatrix} \right]$, 
        i.e., intra-layer edges, $A_{i,j}^{\alpha}$, 
    
    \item \textit{inter-layer graphs}, $G^{\alpha,\beta}$, formed by off-diagonal element blocks, $\left[ \begin{smallmatrix} \boldsymbol{0} & \boldsymbol{A}^{\alpha,\beta} \\ \boldsymbol{A}^{\beta,\alpha} & \boldsymbol{0}  \end{smallmatrix} \right], \alpha \neq \beta$, 
        i.e., inter-layer edges, $A_{i,j}^{\alpha,\beta}$ and $A_{j,i}^{\beta,\alpha}$, and no intra-layer edges, $A_{i,j}^{\alpha} = 0$ and $A_{i,j}^{\beta} = 0$, and 
    
    \item \textit{all-layer graphs}, $G^{\alpha, \beta}_{all}$, formed by both on and off-diagonal element blocks, $\left[ \begin{smallmatrix} \boldsymbol{A}^{\alpha} & \boldsymbol{A}^{\alpha,\beta} \\ \boldsymbol{A}^{\beta,\alpha} & \boldsymbol{A}^{\beta}  \end{smallmatrix} \right], \alpha \neq \beta$, 
        i.e., intra-layer edges, $A_{i,j}^{\alpha}$ and $A_{i,j}^{\beta}$, and inter-layer edges, $A_{i,j}^{\alpha,\beta}$ and $A_{j,i}^{\beta,\alpha}$. 
\end{itemize}
Figure~\ref{fig:mnet} illustrates a simple representation of a multilayer network and the corresponding supra-adjacency matrix. 

Network science encompasses a rich array of methodologies, along with a multitude of topological, statistical, spectral, and combinatorial properties, which are instrumental in the analysis and extraction of information from single-layer networks (see~\cite{Barabasi2016,peach2021hcga}). These methodologies and properties can be seamlessly extended to the structure of MNet and their respective subgraphs. Additionally, there are emerging methods specifically tailored for the study of MNet structures~\cite{kivela2014multilayer}.

\section{Time Series Mappings}\label{sec3} 
The literature presents a wide range of time series mapping methods for converting both UTS and MTS data into network representations. This innovative framework for time series analysis revolves around a mapping function that can draw inspiration from various concepts, including visibility, transition probability, proximity, time series models, and statistical principles~\cite{zou2018complex}.
These mappings can result either in single-layer~\footnote{Recalling the definition of MNet in Section~\ref{subsec:mnet}, a single-layer network, $G,$ is an MNet with $m=1.$} or multilayer networks~\cite{vanessa2020}. 
Until now, the predominant focus has been on strategies for mapping UTS into single-layer networks, while the development of mapping MTS has not been as extensive~\cite{vanessa2020}. 
In particular, the most commonly employed approach involves techniques that condense MTS data into single-layer networks. In this approach, the node set represents the components of the MTS, denoted as $Y_{i,t}$, and the edge set is determined by a statistical model or a correlation measure applied to these UTS components. 
While this method effectively reduces the dimensionality of MTS data into a more compact structure, namely a single-layer network, it comes at the cost of significant data reduction, preserving only the information captured by the models or measures employed within the mapping function~\cite{vanessa2020}.

Mapping MTS to a MNet represents a cutting-edge and promising approach aimed at retaining more comprehensive data information. Initial efforts concentrated on leveraging multiplex networks (see more in~\cite{lacasa2015network,eroglu2018multiplex, vanessa2020}), which constitute a specific type of MNet~\footnote{A multilayer network is considered a multiplex network when $M$ consists of a sequence of $m$ graphs, denoted as ${G_\alpha}{\alpha=1}^m = {(V\alpha, E_\alpha)}_{\alpha=1}^m$, typically sharing the same node set across all layers. Inter-layer edges exclusively connect corresponding nodes in different layers, denoted as $(v_i^{\alpha}, v_j^{\beta}), \alpha \neq \beta, i = j$)~\cite{boccaletti2014structure}.}.
In this approach, each UTS component is mapped onto an individual layer within the multiplex network. Here, every timestamp or its corresponding representation is uniquely represented by a node. Intra-layer edges are established based on the fundamental principles of UTS mapping, while connections between distinct UTS are defined through inter-layer edges linking contemporaneous nodes across consecutive layers. 
A limitation of these mappings lies in the potential loss of lagged cross-correlations during the mapping process, primarily because there are no inter-layer edges connecting lagged nodes directly~\cite{vanessa2023}.

In an attempt to address the aforementioned challenge, our previous work~\cite{vanessa2023} introduced a novel mapping approach based on the concept of visibility. This method maps MTS data to an MNet structure, incorporating intra-layer edges derived from UTS visibility mapping and inter-layer edges connecting lagged nodes using a novel concept known as cross-visibility. 
The results of this proposed mapping approach were promising, demonstrating the MNet's capacity to capture more information compared to both single-layer and multiplex structures. 
Nevertheless, one drawback of this method is its computational intensity, especially when applied to large datasets featuring numerous time series components. This computational load arises from the pairwise computation of cross-visibility between components.

In this work, we introduce a novel mapping approach aimed at mitigating the computational challenge mentioned above. Our primary objective remains to preserve the MNet structure, which encompasses both intra and inter-layer connections. However, we seek to simultaneously reduce data dimensionality and computational complexity. 
To achieve this goal, we turn to Quantile Graphs~\cite{Campanharo2011}.  Building upon the concepts introduced in~\cite{vanessa2023}, we extend the QG methodology to fit within an MNet structure, a topic we delve into in the subsequent section.

\section{\textit{MQG}: a Novel Multivariate Time Series Mapping}\label{sec4}
A Quantile Graph (QG) is the result of a UTS mapping technique rooted in the concept of transition probability, which has consistently demonstrated remarkable efficacy in capturing the essential characteristics of UTS data~\cite{Campanharo2015, Campanharo2017, vanessa2022}. 
This method operates by mapping the serial transition probabilities governing the dynamics between UTS data timestamps using a limited set of symbols, typically sample quantiles.

In this section, we introduce an innovative QG algorithm designed to map MTS data into a Multilayer Quantile Graph (MQG). This algorithm entails the establishment of fresh connections among UTS components. These connections are created by extending the concept of transition probability and incorporating sample quantiles from the UTS components. We start by introducing the QG algorithm tailored for UTS data, followed by the unveiling of the MQG algorithm, customized for MTS data.

\subsection{Quantile Graph}
\label{subsec:qg}
\begin{figure*}[ht]
    \centering
    \includegraphics[width=1\textwidth]{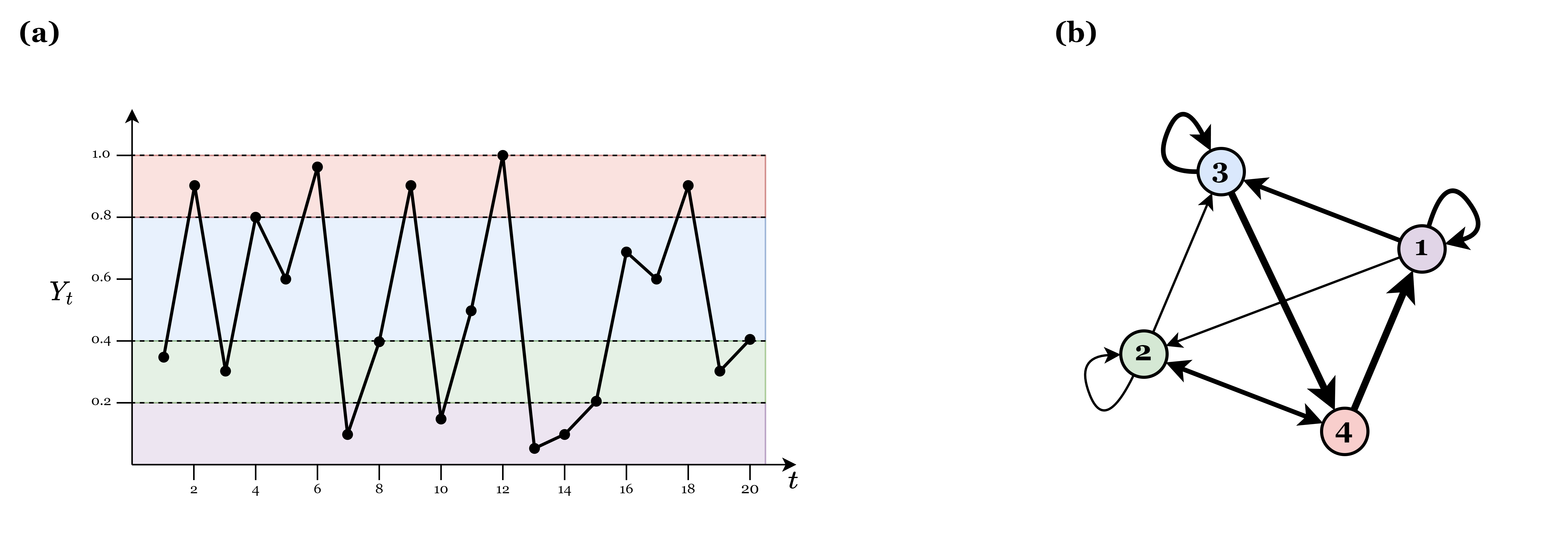}
    \caption[Illustrative example of the Quantile Graph algorithm.]{Illustrative example of the Quantile Graph algorithm for $\eta = 4$. (a) illustrates a toy univariate time series with colored regions representing the different $\eta$ sample quantiles, and (b) the corresponding network generated by the Quantile Graph algorithm. The thicker directed lines represent the edges with greater weights accounting for repeated transitions between quantiles. 
    \textit{Source:} Reproduced from~\cite{vanessa2020}}
    \label{fig:qg}
\end{figure*}

The QG algorithm~\cite{vanessa2020} (see Algorithm~\ref{alg:qg}) starts to assign the UTS observations to bins that are defined by $\eta$ sample quantiles, $q_{1}, q_{2}, ..., q_{\eta}$.  
Each sample quantile, $q_{i}$, is mapped to a node $v_{i}$ of the corresponding graph (a single-layer network) and the edges between two nodes $v_{i}$ and $v_{j}$ are directed and weighted, $(v_{i}, v_{j}, w_{i,j})$, with $w_{i,j}$ corresponding to the transition probability between quantile ranges. 
The adjacency matrix is a Markov transition matrix: $\sum_{j=1}^\eta w_{i,j} = 1$, for each $i = 1, \ldots, \eta,$ and the single-layer network is weighted, directed and contains self-loops\footnote{A self-loop is an edge that connects a node to itself.}. 
Figure~\ref{fig:qg} illustrates this mapping method. 

Typically, the number of quantiles ($\eta$) is significantly smaller than the length of the time series ($\eta \ll T$). 
If $\eta$ is excessively large, the resultant graph may not be connected, resulting in isolated nodes\footnote{An isolated node is a node that is not connected by an edge to any other node.}. Conversely, if $\eta$ is too small, the QG may exhibit a substantial loss of information, characterized by the assignment of high weights to self-loops.

\begin{algorithm}[htb!]
    \caption{Quantile Graph}\label{alg:qg}

\hspace*{\algorithmicindent} \textbf{Input:} {time series $ts$ and number of quantiles $\eta$} \\
\hspace*{\algorithmicindent} \textbf{Output:} {multilayer graph $mnet$ and single layer $layer$ of $mnet$} \\
\hspace*{\algorithmicindent} {\textsc{Which\_GEQ} finds the quantile of a given value}
    
    \begin{algorithmic}[1] 

        \Procedure{QG}{$ts, mnet, layer, \eta$} 

            \State $T \gets ts.size()$ 
            \State $probs \gets \{\}$ 
            \State $q \gets \{\}$ 

            \For{$i \gets 1$ \textbf{to} $\eta$}
                \State $probs[a] \gets i/\eta$\;
            \EndFor

            \State $q \gets $ \Call{Quantiles}{$ts, probs$} 

            \For{$i \gets 1$ \textbf{to} $\eta$}
                \State $mnet.$ \Call{Add\_Node}{$i, layer$} 
            \EndFor

            \For{$i \gets 1$ \textbf{to} $\eta$}
                \State $from \gets $ \Call{Which\_GEQ}{$q, ts[i]$} 
                \State $to \gets $ \Call{Which\_GEQ}{$q, ts[i+1]$} 
                \State $e \gets mnet.$ \Call{Get\_Edge}{$from, to, layer$}  
                
                \If{$!e$}
                    \State $mnet.$ \Call{Add\_Edge}{$from, to, layer, 1$} 
                \Else
                    \State $w \gets mnet.$ \Call{Get\_Weight}{$e$} 
                    \State $mnet.$ \Call{Set\_Weight}{$e, w+1$} 
                \EndIf
                
                \State $layer.q\_seq[i] \gets from$\;
                \If{$i == T-1$}
                    \State $layer.q\_seq[T] \gets to$\;
                \EndIf
            \EndFor

            \State \Return{}

        \EndProcedure

    \end{algorithmic}
\end{algorithm}

\subsection{Multilayer Quantile Graph}
\begin{figure*}[htb!]
    \centering
    \includegraphics[width=0.96\textwidth]{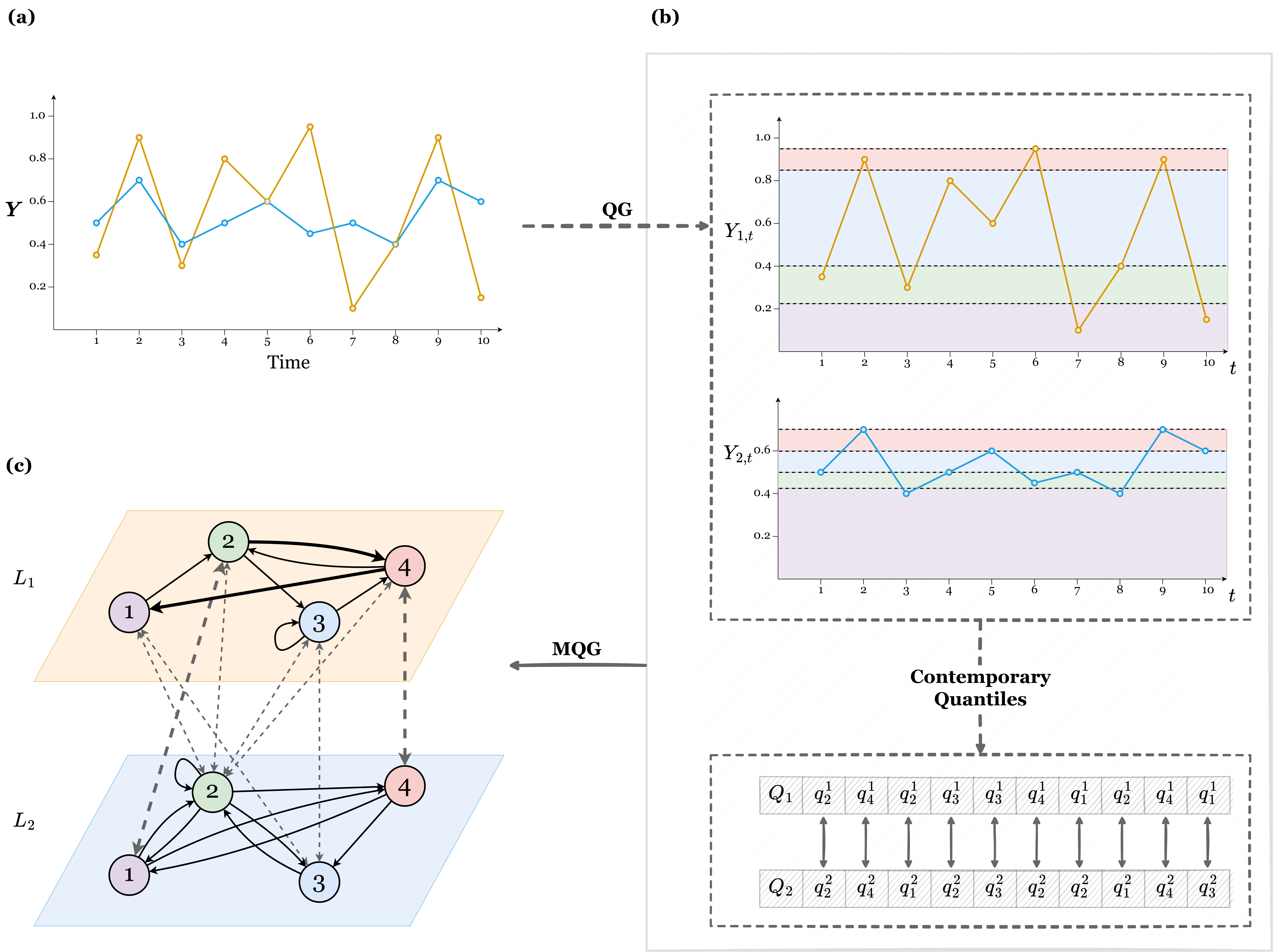}
    \caption[Schematic diagram of the Multilayer Quantile Graph algorithm.]{Schematic diagram of the Multilayer Quantile Graph algorithm for $\eta=4$: 
    (a) original time series, (b) illustration of the intra-layer Quantile Graphs (colored regions representing the different sample quantiles) and inter-layer contemporaneous edges mapping, (d) Multilayer Quantile Graph: black lines represent the intra-layer edges (the QGs), dashed lines the inter-layer edges between nodes contemporaneous quantile nodes, and the thickness of the lines represent the weighted intensities of the edges.}
    \label{fig:mqg}
\end{figure*}

The MQG algorithm (see Algorithm~\ref{alg:mqg}) builds upon the foundational principles introduced in the preceding QG framework. It starts by mapping all UTS components within MTS data into corresponding QGs. Subsequently, for each pair of UTS, denoted as $\boldsymbol{Y}^{\alpha}$ and $\boldsymbol{Y}^{\beta}$, the contemporaneous quantiles ($q_i^{\alpha}$ and $q_j^{\beta}$ at time $t$) are linked by inter-layer edges representing the cross-dimensions contemporary transitions. 
In more detail, the MQG involves the following three steps (see Figure~\ref{fig:mqg}):
\begin{description}
    \item[\textbf{Step 1:}] each UTS component, $\boldsymbol{Y}^{\alpha}, \alpha = 1, \ldots, m,$ is mapped to a QG, $L_{\alpha},$ by applying the Algorithm~\ref{alg:qg} (top of Figure~\ref{fig:mqg}).

    \item[\textbf{Step 2:}] each pair of QGs, $L_{\alpha}$ and $L_{\beta}, \alpha,\beta = 1, \ldots, m$ and $ \alpha \neq \beta,$ is connected by the corresponding contemporary quantiles, $q_i^{\alpha}$ and $q_j^{\beta}, i,j = 1, \ldots, \eta,$ at time $t = 1, \ldots, T$ (panel \textbf{(b)} of Figure~\ref{fig:mqg} and Algorithm~\ref{alg:inter_qg}). 
    The quantiles $q_i^{\alpha}$ and $q_j^{\beta}$ belong to the temporal sequences $\boldsymbol{Q}_{\alpha}$ and $\boldsymbol{Q}_{\beta},$ respectively. 
    The weight $w_{i,j}^{\alpha,\beta}$ represents the probability that $Y_{\alpha,t}$ and $Y_{\beta,t}$ belong to the quantiles $q_{i}^{\alpha}$ and $q_{j}^{\beta},$ respectively, at the same time.

    \item[\textbf{Step 3:}] a MTS, $\boldsymbol{Y},$ is mapped into a MQG, $M,$ (panel \textbf{(c)} of Figure~\ref{fig:mqg}). 
    Layer set refers to each QG, $L_{\alpha} \in \boldsymbol{L}, \alpha=1, \ldots, m,$ and the node-layer set refers to the sample quantiles, $V_M = \{q_i^{\alpha}\}_{i=1}^{\eta}.$ 
    The directed weighted intra-layer edges, $(q_i^{\alpha}, q_j^{\alpha}, w_{i,j}^{\alpha}) \in E_M,$ of individual QG are established in the corresponding layer, $L_{\alpha},$ and the bidirectional weighted inter-layer edges, $(q_i^{\alpha}, q_j^{\beta}, w_{i,j}^{\alpha,\beta}) \in E_M, \alpha \neq \beta,$ of pair-wise QGs are established between dimensions pair-wise layers, $L_{\alpha}$ and $L_{\beta}.$ 
\end{description}

\begin{algorithm}[htb!]
    \caption{Multilayer Quantile Graph}
    \label{alg:mqg}

\hspace*{\algorithmicindent} \textbf{Input:} {multivariate time series $mts$ and number of quantiles $\eta$} \\
\hspace*{\algorithmicindent} \textbf{Output:} {multilayer graph $mnet$}
    
    \begin{algorithmic}[1] 

        \Procedure{MQG}{$mts,\eta$} 

            \State $m \gets mts.size()$ 
            \State $mnet \gets \{\}$ 

            \For{$a \gets 1$ \textbf{to} $m$}
                \State $mnet.layers[a] \gets \{\}$ 
                \State \Call{Set\_Direction}{$mnet.layers[a], true$} 
                
                \State \Call{QG}{$mts[a], mnet, mnet.layers[a], \eta$} 
            \EndFor

            \For{$a \gets 1$ \textbf{to} $m-1$}
                \For{$b \gets a+1$ \textbf{to} $m$}
                    \State \Call{Set\_Direction}{$mnet.layers[a], mnet.layers[b], false$} 
                    \State \Call{Contem\_QG}{$mnet, mnet.layers[a], mnet.layers[b]$} 
                \EndFor
            \EndFor

            \State \Return $mnet$

        \EndProcedure

    \end{algorithmic}
\end{algorithm}

\begin{algorithm}[htb!]
    \caption{Contemporaneous Quantile Graph}
    \label{alg:inter_qg}

\hspace*{\algorithmicindent} \textbf{Input:} {multilayer network $mnet$} \\
\hspace*{\algorithmicindent} \textbf{Output:} {two specified layers ($layerA$ and $layerB$) of $mnet$} 
    
    \begin{algorithmic}[1] 

        \Procedure{Contem\_QG}{$mnet, layerA, layerB$} 

            \State $T \gets layerA.size()$ 
            \State $qA \gets layerA.q\_seq$ 
            \State $qB \gets layerB.q\_seq$ 

            \For{$i \gets 1$ \textbf{to} $T$}
                \State $e \gets mnet.$ \Call{Get\_Edge}{$qA[i], qB[i], layerA, layerB$} 
                \If{$!e$}
                    \State $mnet.$ \Call{Add\_Edge}{$qA[i], qB[i], layerA, layerB, 1$} 
                \Else
                    \State $w \gets mnet.$ \Call{Get\_Weight}{$e$} 
                    \State $mnet.$ \Call{Set\_Weight}{$e, w+1$} 
                \EndIf
            \EndFor

            \State \Return{}

        \EndProcedure

    \end{algorithmic}
\end{algorithm}

\newpage
MQG is a directed and weighted MNet. Note that the inter-layer edges are bi-directional, that is, whenever there is a transition from $q_i^{\alpha}$ to $q_j^{\beta}$ there is an equivalent transition from $q_j^{\beta}$ to $q_i^{\alpha}.$ This symmetry implies that inter-layer edges can also be equivalently represented as undirected edges.
To clarify the relationship with the MNet subgraphs described in Section~\ref{subsec:mnet}, we can distinguish two key components within the MQG: a) intra-layer graphs, which correspond to individual QGs, and; b) inter-layer graphs, which correspond to bipartite graphs representing contemporaneous transitions. We will refer to these bipartite graphs as Contemporaneous Quantile Graphs.

The version of the MQG algorithm presented here is the basic version that maps the interconnections contemporaneously, however, the algorithm can be extended to represent transitions between quantiles corresponding to lagged timestamps from two different layers (rather than just between consecutive quantiles). 

\section{Analyzing Synthetic Data Set via MQG}\label{sec5}
Analyzing time series data by leveraging features extracted directly and indirectly from the data has emerged as a recent and promising approach in time series data mining~\cite{Wang2006,henderson2021empirical,vanessa2022}.  
In this section, we will use the topological features introduced in~\cite{vanessa2023}. These features are based on both intra-layer and inter-layer edges, and our goal is to apply them to analyze the proposed MQG method. The aim is to ascertain whether the information conveyed by inter-layer edges complements the insights gained from intra-layer edges, which is the conventional approach in the literature.

Much like the approach adopted in~\cite{vanessa2023}, we employ a rich synthetic MTS dataset and employ data mining techniques to analyze this dataset.

\subsection{Data set: Multivariate Time Series Models}
We use the six linear and nonlinear bivariate time series models ($m=2$) summarized in Table~\ref{tab:mts_setting} and described in detail in~\cite{vanessa2023}. 
For each of the six different MTS models, we have a total of 100 instances of length $T = 10000$, and the parameters are chosen so that the data exhibits a range of serial and cross-correlation properties as described in Table~\ref{tab:mts_setting}. 
From here on, we refer to this data set as Multivariate Data Generating Processes (MDGP). MDGP is available at~\url{https://github.com/vanessa-silva/MHVG2MTS}.

\begin{table*}[ht]
\centering
    \caption{Summary of synthetic bivariate time series processes: parameters, the main characteristic of the data, and notation. See~\cite{vanessa2023} for more details.}
    \label{tab:mts_setting}

\begin{tabular}{@{}lllc@{}}
\toprule
\textbf{MDGP} & \textbf{Parameters} & \textbf{Characteristics} & \textbf{Notation} \\
\midrule

\rule{0pt}{12pt}Independent White Noise & $\displaystyle \boldsymbol{\epsilon}_t \sim N(0,1)$ & \multirow{2}{*}{\begin{tabular}{@{}l@{}}Noise effect \\ No correlation\end{tabular}}  &  \texttt{iBWN} \\
 & &  & \\[1ex]
\rule{0pt}{12pt}Correlated White Noise & $\displaystyle \bigl[ \begin{smallmatrix} \epsilon_{1,t} \\ \epsilon_{2,t} \end{smallmatrix} \bigr] \sim N\left(0, \bigl[ \begin{smallmatrix} 1.00 & 0.86\\ 0.86 & 1.50 \end{smallmatrix} \bigr] \right)$ &  \multirow{3}{*}{\begin{tabular}{@{}l@{}}Noise effect \\ No serial correlation \\ Cross-correlation \end{tabular}}  &  \texttt{cBWN} \\
& &   & \\
& &   & \\[1ex]

\rule{0pt}{12pt}Weak VAR$(1)$ & $\displaystyle \boldsymbol{\varphi} = \bigl[ \begin{smallmatrix} 2.50 \\ 0.50 \end{smallmatrix} \bigr], \boldsymbol{\phi} = \bigl[ \begin{smallmatrix} 0.20 & 0.10 \\ 0.02 & 0.10 \end{smallmatrix} \bigr]$ & \multirow{2}{*}{\begin{tabular}{@{}l@{}}Weak correlation \\ (serial and cross)\end{tabular}}  &  \texttt{wVAR} \\[1ex]
& $\displaystyle \boldsymbol{\epsilon}_t \sim \bigl[ \begin{smallmatrix} 1.00 & 0.10 \\ 0.10 & 1.50 \end{smallmatrix} \bigr]$ &  & \\[1ex]
\rule{0pt}{12pt}Strong VAR$(1)$ & $\displaystyle \boldsymbol{\varphi} = \bigl[ \begin{smallmatrix} 0 \\ 0 \end{smallmatrix} \bigr], \boldsymbol{\phi} = \bigl[ \begin{smallmatrix} 0.70 & 0.02 \\ 0.30 & 0.80 \end{smallmatrix} \bigr]$ & \multirow{3}{*}{\begin{tabular}{@{}l@{}}Strong correlation \\ (serial and cross, lagged \\ and contemporaneous)\end{tabular}}  &  \texttt{sVAR} \\[1ex]
 & $\displaystyle \boldsymbol{\epsilon}_t \sim \bigl[ \begin{smallmatrix} 1.00 & 0.86 \\ 0.86 & 1.50 \end{smallmatrix} \bigr]$ &  & \\
  & &  \\[1ex]

\rule{0pt}{12pt}Weak VGARCH$(1,1)$ & $\displaystyle\boldsymbol{\omega} = \bigl[ \begin{smallmatrix} 0.05 \\ 0.02 \end{smallmatrix} \bigr]$, $\boldsymbol{\alpha} = \bigl[ \begin{smallmatrix} 0.10 & 0.00 \\ 0.00 & 0.05 \end{smallmatrix} \bigr] $ & \multirow{2}{*}{\begin{tabular}{@{}l@{}}No serial correlation \\ Weak cross-correlation\end{tabular}} &  \texttt{wVGARCH} \\[1ex]
& $\displaystyle \boldsymbol{\beta} = \bigl[ \begin{smallmatrix} 0.85 & 0.00 \\ 0.00 & 0.88 \end{smallmatrix} \bigr], \boldsymbol{\epsilon}_t \sim \bigl[ \begin{smallmatrix} 1.00 & 0.10 \\ 0.10 & 1.50 \end{smallmatrix} \bigr]$ &  & \\[1ex]
\rule{0pt}{12pt}Strong VGARCH$(1,1)$ & $\displaystyle \boldsymbol{\omega} = \bigl[ \begin{smallmatrix} 0.05 \\ 0.02 \end{smallmatrix} \bigr]$, $\boldsymbol{\alpha} = \bigl[ \begin{smallmatrix} 0.10 & 0.00 \\ 0.00 & 0.05 \end{smallmatrix} \bigr] $ & \multirow{2}{*}{\begin{tabular}{@{}l@{}}Strong contemporaneous \\ cross-correlation\end{tabular}} &  \texttt{sVGARCH} \\[1ex]
 & $\displaystyle \boldsymbol{\beta} = \bigl[ \begin{smallmatrix} 0.85 & 0.00 \\ 0.00 & 0.88 \end{smallmatrix} \bigr], \boldsymbol{\epsilon}_t \sim \bigl[ \begin{smallmatrix} 1.00 & 0.86 \\ 0.86 & 1.50 \end{smallmatrix} \bigr]$ &   & \\[1ex]

\end{tabular}
\end{table*}

In short, MDGP is a diverse set of MTS models with a specific set of properties related to serial and cross-correlation characteristics, namely:  
    \textit{white noise} (WN) processes representing the noise effects, 
    \textit{vector autoregression} (VAR) processes representing smooth linear data, and 
    \textit{vector generalized autoregressive conditional heteroskedasticity} (VGARCH) processes representing nonlinear data with high or low volatility.
    Furthermore, these processes are designed to represent different levels of correlation, that is, data with and without serial and cross-correlation and data with weak and strong correlation properties both serial and cross, as well as lagged and contemporaneously. 
A detailed description of the MDGP and their properties, as well as computational details, can be seen in~\cite{vanessa2023}. 

\subsection{Implementation Details}
We start by mapping each bivariate time series model into the corresponding MQG using the Algorithm~\ref{alg:mqg} presented in Section~\ref{sec4}. 
We use $\eta = 50$ quantiles, as in~\cite{campanharo2016hurst,vanessa2022}, which is also a value close to the result of the formula $\eta \approx 2T^{1/3}$ defined in~\cite{campanharo2018application} that relates the number of quantiles to the time series length. 
We also highlight the intra-, inter-, and all-layer graphs using the corresponding adjacency submatrices to the MQG as defined in Section~\ref{subsec:mnet}. 

To analyze the MDGP and evaluate the proposed MQG method, we extract corresponding topological features from each resulting MQG. 
For this, we use the high-dimensional topological features presented in~\cite{vanessa2023} and use the methodologies and algorithms described in this work. 
In short, we compute the following features for the resulting MQGs and its subgraphs: 
\begin{itemize}
    \item \textbf{average degree ($\bar{k}$)}: computing the arithmetic mean of the degrees $k_i$ of all node $v_i$ in the respective subgraph;

    \item \textbf{average path length ($\bar{d}$)}: using an algorithm that computes the average shortest path length between all pairs of nodes (of respective subgraphs) using a breadth-first search algorithm;

    \item \textbf{modularity ($Q$)}: computing how good a specific division of the corresponding subgraph into communities is, based on the number of triangles and number of triples;
    
    \item \textbf{number of communities ($S$)}: using a function that makes use of the known "Louvain" algorithm that finds community structures by multi-level optimization of \textbf{modularity ($Q$)} feature (see~\cite{blondel2008fast} for more details),

    \item \textbf{average ratio degree ($\bar{r}$)}: computing the arithmetic mean of the ratio degrees $r_i$ of all node $v_i$ in the respective subgraph (this is a new topological measure proposed in~\cite{vanessa2023});
    
    \item \textbf{Jensen–Shannon divergence ($JSD$)}: computing the similarity between two degree distributions using the known Jensen–Shannon divergence measure.
\end{itemize}
We calculate the first four measures in the three different possible subgraphs, i.e., intra-, inter-, and all-layer graphs, and the resulting set of features are called \textit{intra-features}, \textit{inter-features} and \textit{all-features}. 
The last two measures are called \textit{relational features}, as they either measure the similarity/relationship between different connections or measures. 
This results in a unique vector of 21 features.

We used C++ and its needed set of libraries (such as \texttt{igraph} and standard libraries) to implement the data structure to store an MNet and compute the functions to extract the topological features.

\subsection{Mining MDGP with MQG Features}

\begin{figure*}[htb!]
    \centering
    \includegraphics[width=0.75\textwidth]{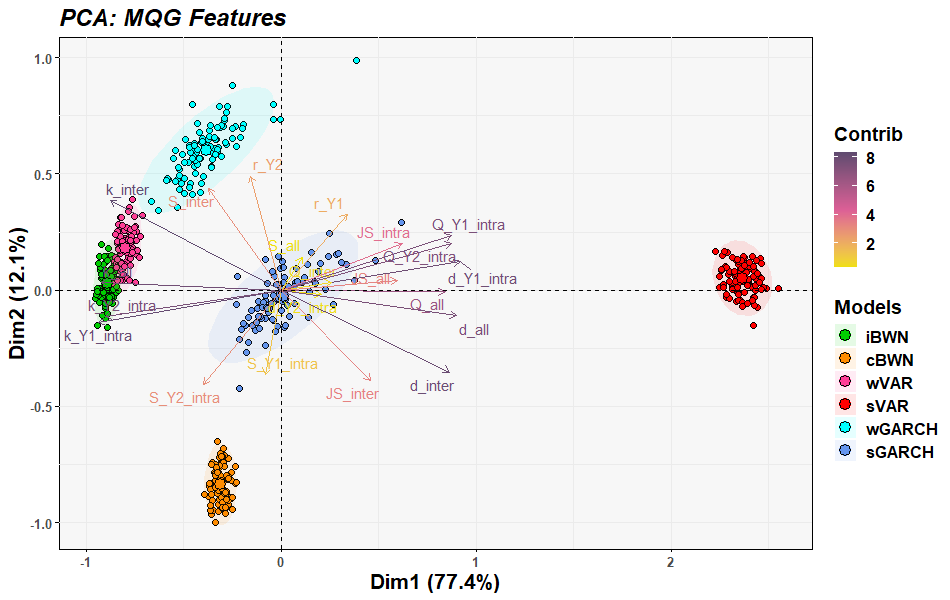}
    \caption{Bi-plot of the first two PCs of MQG topological feature set for the synthetic bivariate dataset. Different colors represent the different Multivariate Data Generating Processes and the arrows represent the contribution of the corresponding feature to the PCs:  the larger the size, the sharper the color, and the closer to the red the greater the contribution of the feature. 
	Features grouped are positively correlated while those placed on opposite quadrants are negatively correlated.}
	\label{fig:pca_mqg}
\end{figure*}

From the resulting diversified vector of 21 high-dimensional topological features (intra-, inter-, all-layer, and relational features) extracted from the proposed multilayer time series network, MQG, we perform a principal component analysis (PCA). 
The obtained PCA feature space is illustrated in Figure~\ref{fig:pca_mqg} and shows which MNet topological features capture the different properties of the MDGP. 

The feature space is shown in a bi-plot obtained using a total of 21 features with the two PCs explaining 89.5\% of the data variance. 
We can see a good distribution of the characteristics inherent to each sample model. 
All topological features contribute to the arrangement of the samples which capture different properties of the MTS models. 
In particular, we can see that the average degree and the number of communities of intra-layer graphs of MQG try to place the WN models in the third quadrant while the same features for inter-layer graphs try to place the VGARCH and VAR models weakly correlated in the second quadrant. The communities-based features and the average ratio degree seem to contribute to distinguishing the strong and the weak correlation of heteroskedastic models.
 
Additionally, we evaluate the MQG feature set (vector of 21 MNet features) in a case study regarding time series clustering. 
For this task, we rescale the intended topological feature vector into the [0, 1] interval using Min-Max normalization, the PCs are computed (no need of z-score normalization within PCA) and a clustering algorithm, \textit{k}-means, is applied to the PC’s corresponding to 100\% of variance. 
The clustering results are assessed using appropriate evaluation metrics: \textit{Average Silhouette} (AS); \textit{Adjusted Rand Index} (ARI) and \textit{Normalized Mutual Information} (NMI). Note that AS does not need ground truth, while ARI and NMI do. The range of values for NMI is [0, 1] and for ARI and AS is [-1, 1].

We start by analyzing the usefulness of the MQG feature set by performing a clustering exercise considering different subsets of the MQG feature set. 
The results are summarized in Table~\ref{table:clust_res_subset} and indicate that inter-layer edges contain additional information about the MTS data, leading to better clustering results (together with the intra-layer edges). 
We can also analyze that relational features achieve good clustering results when considered alone. 
Sub-graphs with both intra-layer edges and inter-layer edges add information that leads to improvements in the clustering results (compare the last three rows of  Table~\ref{table:clust_res_subset} with the first two). The results show that the Contemporaneous Quantile Graphs (the inter-layer edges from MQG) capture different properties from MTS. 
Note that the results from the set of intra-layer features are good because the MDGP under analysis involves the same statistical process for the two components of time series whose properties inherent to each process are also captured by the QG mapping methods (as we see in~\cite{vanessa2022}). 

\begin{table}[h!]
\centering
\caption{Clustering evaluation metrics for the different clustering analyses resulting from different MQG feature vectors.   
The values reflect the mean of 10 repetitions of the proposed method for different feature vectors and the ground truth ($k = 6$). 
The two highest values are highlighted in each column.}

\begin{tabular}{|l|c|c|c|}
\hline

\multicolumn{1}{|c|}{\multirow{2}{*}{\textbf{Feature Set}}} 
    & \multicolumn{1}{c|}{\textbf{ARI}} & \multicolumn{1}{c|}{\textbf{NMI}} & \multicolumn{1}{c|}{\textbf{AS}} \\
    & \multicolumn{1}{c|}{\scriptsize $[-1,1]$} & \multicolumn{1}{c|}{\scriptsize $[0,1]$} & \multicolumn{1}{c|}{\scriptsize $[-1,1]$} \\

\hline
\textbf{\textit{Intra}-layer} 		 &  0.52              &       0.65              &        0.52      \\
\hline
\textbf{\textit{Inter}-layer} 		 &      0.57             &       0.68              &  \textbf{0.83}  \\
\hline
\textbf{\textit{All}-layer} 		    &   \textbf{0.78}      &   \textbf{0.86}    &  \textbf{0.78} \\
\hline
\textbf{\textit{Relational}}         & {0.71}  & {0.81}   &   {0.55}  \\
\hline

\hline
\textbf{\textit{MQG}}    		    
  & \textbf{0.96}     & \textbf{0.96}     & {0.53} \\
\hline

\end{tabular}
\label{table:clust_res_subset}
\end{table}

We also compare the results obtained using the MQG feature set with the results obtained in~\cite{vanessa2023}, that is, using the same feature set but for the MHVG mapping method. 
Table~\ref{table:clust_res} summarizes the clustering results obtained in~\cite{vanessa2023}. The two experiments are made in the same computational environment and using the same methods. 
We can conclude that features from MQGs are more accurate, almost perfect when we look at the evaluation features ARI and NMI, with a mean value of 0.96, and AS with 0.53, when compared to cross-visibility based mapping that obtains 0.63 to ARI, 0.71 to NMI, and 0.45 to AS feature. 
So, for the MDGP used in this work, the MQG can be sufficient to cluster the different MTS model samples. 

\begin{table}[h!]
\centering

\caption{Clustering evaluation metrics for the different clustering analyses resulting from different MHVG feature vectors.   
The values reflect the mean of 10 repetitions of the proposed method for different feature vectors and the ground truth ($k = 6$). 
The two highest values are highlighted in each column. 
\textit{Source}: Extracted from~\cite{vanessa2023}.}

\begin{tabular}{|l|r|r|r|}
\hline
\multicolumn{1}{|c|}{\multirow{2}{*}{\textbf{Feature Set}}} 
    & \multicolumn{1}{c|}{\textbf{ARI}} & \multicolumn{1}{c|}{\textbf{NMI}} & \multicolumn{1}{c|}{\textbf{AS}} \\
     & \multicolumn{1}{c|}{\scriptsize $[-1,1]$} & \multicolumn{1}{c|}{\scriptsize $[0,1]$} & \multicolumn{1}{c|}{\scriptsize $[-1,1]$} \\
  
\hline
\textbf{\textit{Intra}-layer} 					& 0.52             & 0.61             & 0.29             \\
\hline
\textbf{\textit{Inter}-layer} 					& 0.29             & 0.42             & \bf{0.51} \\
\hline
\textbf{\textit{All}-layer} 					& \textbf{0.67}    & \textbf{0.73}    & \bf{0.51} \\
\hline
\textbf{\textit{Relational}}    		        & {0.58} & {0.65} & \textbf{0.62}    \\
\hline
\hline
\textbf{\textit{MHVG}}    		                & \textbf{0.63}    & \textbf{0.71}    & 0.45             \\
\hline
\end{tabular}
\label{table:clust_res}
\end{table}

\begin{figure}
    \centering
    \includegraphics[width=0.55\textwidth]{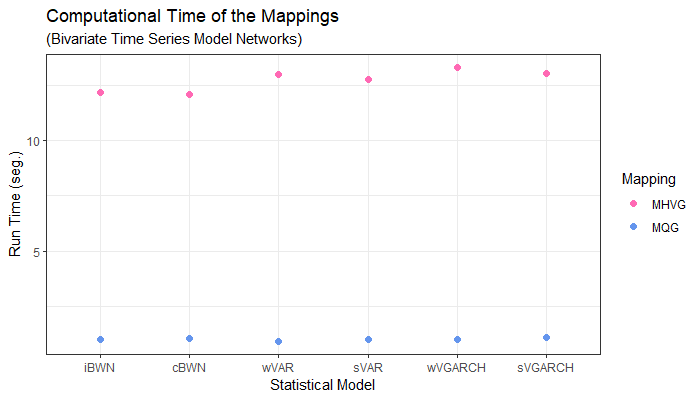}
    \caption{Comparison of the runtime (in seconds) of the MQG and MHVG mapping algorithms for mapping the six different sets of bivariate time series models onto the respective multilayer networks. Each record represents the runtime of just the mapping method of the 100 instances of each model in the respective MQG or MHVG.}
	\label{rtime}
\end{figure}

The big advantage of MQG mapping over MHVG mapping is that MQG is a significantly reduced representation of the original MTS data set, which makes the mapping method in this case consistently an order of magnitude faster for all models, as we can see in Figure~\ref{rtime}. 
Runtimes presented in Figure~\ref{rtime} are the times corresponding only to the execution of the mapping methods (MQG or MHVG). Note that the difference would be even greater given the fact that it would then still be necessary to compute the topological features in the resulting multilayer networks. 
Furthermore, for very large datasets the MQG method is executable, whereas the MHVG method may be impractical.
In addition to the reduction in data dimensionality, our exploratory analysis allows us to analyze that MQG can capture the dynamic properties of the original data, both serial and cross-sectional properties.

\section{Conclusions} \label{sec6}

Methods for mapping multivariate time series into multilayer networks have attracted the attention of researchers in recent times due to their potential. Special interest lies in mappings that generate inter-layer edges thus allowing to capture not only the time dependencies but also the dependencies between different variables.

In this work, we introduce a Multilayer Quantile Graph as a new multivariate time series mapping method. 
MQG is based on a transition probability concept extending the traditional concept of Quantile Graph~\cite{Campanharo2011} for the univariate time series data. 
The procedure consists of two steps. 
    First, create a reduced multilayer network structure that represents a high-dimensional MTS following the QG mapping method available in the literature. The resulting set of layers/graphs characterizes the serial dynamic transitions of each of the time series components.
    Second, for each pair of time series components in the MTS, introduce weighted inter-layer edges between corresponding layers that capture the contemporaneous (and lagged) dynamic transitions between the different time series dimensions.
MQG was designed to reduce the dimensionality of time series data with high dimensionality and be more computationally efficient and feasible than recently proposed mapping methods.  
The resulting multilayer networks have smaller dimensions than 
those obtained by other mapping methods such as the MHVG, and allow the effective reduction of the dimensionality of the MTS. 
To analyze the proposed multivariate time series mapping, MQG, we consider the specific set of multidimensional topological features proposed by~\cite{vanessa2023} for MQGs.  
These features are based on conventional concepts of node centrality, graph distances, clustering, communities, and similarity measures, and are extracted from all the subgraphs of the resulting multilayer network, that is, intra-layer graphs, inter-layer graphs, and all-layer graphs.

To assess the proposed methodology we use the set of MTS models presented in~\cite{vanessa2023}. 
The data set consists of 600 synthetic bivariate time series grouped into six different multivariate statistical models. 
We map the MTS into the MQG and compute the corresponding topological features.  
The analysis of the set of topological features on the feature space provided by the two principal components shows that different topological features (based on different concepts and different subgraphs of the multilayer network) capture different dynamic properties of the time series models. 
Furthermore, comparing the feature spaces obtained from MHVG (see~\cite{vanessa2023}) and from MQG, we can say that the latter enhances the capture of cross-correlation properties. 

Finally, we performed a clustering analysis of the synthetic time series based on topological features obtained from MQG and MHVG.
The results show that, despite the dimensionality reduction, the MQG mapping is sufficient to distinguish the characteristics inherent to each statistical model analyzed in this work and that the inter-layer edges add valuable information to intra-layer features, improving the accuracy of results. 

To conclude, the proposed MQG reduces the dimensionality of the original time series data, reducing the amount of data observations to a smaller number of sample quantiles, preserving the dynamic characteristics of the time series (serially) and between time series (crossly), using probability transitions, during the mapping process. 
The objectives inherent to the design of this mapping method are quite relevant at a multidisciplinary level where the capabilities of the resulting networks are promising. 
In this work, we address cross transitions between quantiles corresponding to consecutive timestamps, but the proposed method can be modified to map cross transitions between quantiles corresponding to lagged timestamps. 
This version can enrich the results obtained, just as in the univariate case~\cite{campanharo2016hurst}.
In our future work, we intend to explore this MQG version and a more detailed analysis in real-world scenarios.

\section*{Availability of data and materials}

The raw data are available from the corresponding author upon request.

\section{Acknowledgments}

This work is financed by National Funds through the Portuguese funding agency, FCT - Fundação para a Ciência e a Tecnologia, within project LA/P/0063/2020.

\bibliographystyle{apalike}
\bibliography{refs}

\newpage

\appendix

\end{document}